\documentstyle[12pt]{article}
\newcommand{\tr}[1]{\,{\rm tr}\,#1\,}

\begin{document}

\title{
\begin{flushright}
{\small SMI-1-93 \\ January, 1993 }
\end{flushright}
\vspace{2cm}
A Simple Model Inducing QCD}
\author{I.Ya. Aref'eva \thanks{E-MAIL: Arefeva@qft.mian.su}
\\ Steklov Mathematical Institute,\\ Russian Academy of Sciences,\\
Vavilov st.42, GSP-1,117966, \\ Moscow, Russia }
\date{~}
\maketitle
\begin {abstract}
A simple lattice model  inducing  a  gauge  theory is considered. The model
describes an interaction of a gauge field to an $N\times N$ complex matrix
scalar  field  transforming as a field in the
fundamental representation. In contrast to the Kazakov-Migdal
model the model contains only the linear interaction between scalar and
gauge lattice fields. This model does not suffer from extra local $U(1)$
symmetries.

In an approximation of a translation invariant master field the large N limit
of the model is investigated. At large N the gauge fields can be integrated out
yielding an effective theory describing an interaction of eigenvalues of the
master field. The reduced model exhibits phase transitions at the points
$\beta _{\bar {cr}}$ and
$\beta _{\underline{cr}}$  and the region $(\beta _{\bar{cr}},
\beta _{\underline{cr}})$ separates the strong and weak regions of the model.

To study the behaviour of the model at large $N$ in more systematic way
the quenched momentum  prescription
with constraints for treating the large N limit of gauge theories
is used. With the help of the technique of orthogonal
polynomials   nonlinear equations describing the large N limit of the
reduced model {\it with quenching} are presented.
\end {abstract}
\newpage
\section{Introduction}
This paper is inspired by recent studies of the  Kazakov  Migdal (KM) model
\cite {KM} - \cite {D'AT}, which was suggested as a possible model of hadrons.
The KM model contains a scalar field $\Phi$ in the adjoint representation of
the $SU(N)$ group interacting with the usual lattice gauge field. The main
point
which allows to study the large N limit of the model analytically, is the
absence of the kinetic term for the gauge field.

There are several problems with the KM model.
A relation of this model with QCD and, in particular, the property
of asymptotic freedom still should be clarified.
The KM model has  an extra local gauge symmetry  \cite {KMSW,Gross}, which
should be broken if we want
to recover QCD from this lattice model.  There is also a more technical
problem,
concerning the method of investigation of the large N limit in the model.
Most of previous considerations used the approximation of the constant master
field. To clarify this assumption  the large N limit of the KM model has been
investigated by an other method, namely, by the quenched momentum  prescription
 \cite {EK} -  \cite {Das}
for study the large N limit of gauge theories \cite {IA92}. It has been
shown that using the quenched procedure with constraints one obtains the one
side model with the quartic interaction, that makes the model
not exactly solvable.

In this  paper we are going to  construct a modified model which does not
suffer
from extra gauge symmetry and admits more simple analytical treating
at large $N$.

Several modifications of the KM model have been proposed recently.
Migdal \cite {MigMM}, and Khokhlachev and Makeenko  \cite {KhMAF}
proposed to induce QCD by fermions in the adjoint representation of the gauge
group $SU(N)$ with various types of lattice fermions. They discussed also
adjoint scalar and fermionic models at large number of flavours $N_{f}$.

We will consider a simple lattice  gauge model in which  an interaction
of gauge fields  is induced  by a  scalar  field, being an $N\times N$
complex matrix transforming as a field in the
fundamental representation. In contrast to the KM model the modified model
contains only {\it linear} interaction between scalar and
gauge lattice fields. This model does not suffer from extra local $U(1)$
symmetries.  To estimate the behaviour of this model we perform
an integration over lattice gauge fields and then use a translation
invariant master field to calculate the remaining integral over scalar field.
The integration over gauge fields at large N is performed by using the
Brezin-Gross results \cite {BG} of study one link gauge model in an external
matrix field.  The phase transition which
takes place at large N  for one link gauge model in an external matrix field
 \cite {GW,BG} gives rise to a non-trivial phase structure of our model.
The reduced model exhibits   phase transitions at the points
$\beta _{ \bar {cr}}$  and $\beta _{\underline{cr}}$  and the region
$(\beta _{\bar{cr}},\beta _{\underline{cr}})$ separates the strong and weak
coupling regimes of the model.

To study the behaviour of the model at large $N$ in a more systematic way
the quenched momentum  prescription
with constraints for treating the large N limit of gauge theories
is used.  This prescription   leads to a quadratic
dependence of the reduced action on unitary matrix $V$  instead of a quartic
dependence for the KM model. This quadratic dependence permits to use the
technique of orthogonal polynomials  \cite {Metha,KM} and to obtain
nonlinear equations describing the large N limit of the
reduced model  {\it with quenching}. These equations are similar to equations
describing the KM  reduced  model {\it without quenching}.

\section   {The Action of the Model}
Let us consider a simple lattice model which describes  an interaction between
$N\times N$ matrix scalar field $\Phi$ and gauge field and $U_{\mu}$. The
partition
function is given by
\begin {equation} 
                                                          \label {1.1}
Z=\int \prod_{x,\mu} dU_{\mu}(x) \prod_x d\Phi(x)
\exp \{S(\Phi(x), U_\mu(x))\}.
\end   {equation} 
Where $\Phi$ is an arbitrary $N\times N$ matrix and the scalar field $\Phi (x)$
defined on the sides of a d-dimensional
rectangular lattice  and $U_{\mu}$ is an unitary matrix defined on the links;
$d\Phi =\prod _{i,j}d\phi_{ij}$  and $dU_{\mu}$ is the Haar measure on
the group of unitary matrices. The action has the following  form
\begin {equation} 
                                                          \label {1.2}
S(\Phi, U)=-\frac{1}{2}m^{2}N\tr \Phi \Phi^\dagger +\beta N \sum _{\mu >0}
\tr [\Phi (x)U_{\mu}(x)\Phi ^\dagger (x+\mu)+ \Phi (x+\mu)U_{\mu}^\dagger (x)
\Phi ^\dagger (x)].
\end   {equation} 
As in the case of the KM model in (\ref {1.2}) there is no a kinetic term
for the gauge field $U_{\mu}$.

The action (\ref {1.2}) is invariant
under gauge transformations
\begin {equation} 
                                                          \label {1.3}
U_{\mu}(x)\to \Omega (x)U_{\mu}(x)\Omega ^\dagger (x+\mu),~~
\Phi (x) \to \Phi (x)\Omega ^\dagger (x), ~~
 \Phi (x)^\dagger \to \Omega (x)\Phi ^\dagger(x)
\end   {equation} 

Note that one can consider the matrix field $\phi _{ij}$ as a field with
upper and down indices $\phi _{ij}=\phi _{i}^{j}$  and regard the upper
index as the usual colour index and the down as the flavour index. So, with
the respect to the down indeces the scalar field transforms as the usual matter
field in the fundamental representation, and the action is the sum of N
copies of matter actions in the fundamental representation.
In accordance with this analogy the action is invariant under the global $U(N)$
symmetry
\begin {equation} 
                                                          \label {1.4}
\Phi (x) \to  G ^\dagger \Phi (x), ~~
 \Phi (x)^\dagger \to \Phi ^\dagger(x) G,
\end   {equation} 
$G$ belongs to $U(N)$. Note  also that in contrast to the KM model the  model
(\ref {1.2}) does not
have  extra {\it local } $U(1)$ symmetries.

Calculating gaussian integral over fields $\Phi$ and $\Phi ^\dagger$  one
rewrites
the model (\ref {1.2}) as a pure gauge theory with the following reduced action
\begin {equation} 
                                                          \label {1.5}
S_{ind}=N\sum _{\Gamma}(\frac{m^{2}_{0}}{\beta})^{-l(\Gamma)}
\tr U(\Gamma)
\end   {equation} 
Notice that the expansion (\ref {1.5})  starts with the standard Wilson action.
Recall that the analogous expansion for the KM model starts  with the square
of the Wilson action. To get the  Wilson action  instead of the action
(\ref {1.5}) one has to find a mechanism of suppression of long loops in the
effective action (\ref {1.5}). For example, one can use the following trick,
which  was also used by Khokhlachev and Makeenko \cite {KhMAF} to the KM model.
Let attache an extra index to the field
$\Phi$, i.e. introduce the field $\Phi ^{f}$, $f=1,...n_{f} $ and
consider $m_{0}\to \infty$ and
$n_{f}\to \infty$ so that $m_{0}^{2}\sim (n_{f})^{1/4}$.
Then  only  the single-plaquette survives in (\ref {1.5}), the longer loops are
suppressed at least as $n_{f}^{-1/2}$.

A naive continuous version of our model corresponds to $N$ copies of the gauged
scalar fields in the fundamental representation. These N copies can be
interpret
as fields with different flavour and $N_{c}=N_{f}$. The corresponding
Lagrangian has
no an explicit Yang-Mills term
\begin {equation} 
                                                          \label {1.6}
{\cal L}=\frac{N}{g_{0}}(\partial \Phi-i\Phi A_{\mu})(\partial \Phi ^\dagger )
+i A_{\mu}\Phi ^\dagger  +V(\Phi \Phi  ^\dagger).
\end   {equation} 
Here there is not the problem which arises in the naive continuous version of
the KM model, where a commutator $[A_{\mu}, \Phi]$  presents and therefore we
miss the diagonal degrees of freedom of the gauge field.

\section   {Reduced Model}

In this section we estimate the partition function (\ref {1.1})
in the large N limit  using the quenched momentum prescription  \cite {EK} -
\cite {Das}.
This prescription for treating  the limit of infinite N
 of matrix field theories has been obtained more than
ten years ego by Eguchi and Kawai \cite {EK}, Bhanot, Heller and Neuberger
 \cite {BHN},  Parisi  \cite {Parisi}, and
Gross and Kitazawa  \cite {GK} (see also  \cite {IA82} - \cite {Das}).
Generally speaking, according to this prescription
one gets the vacuum energy in the large N limit by integration the free
energy of the quenched reduced model over the quenching parameter $p$. The
reduced
model is described by a reduced action and to get the reduced action one should
replace
a matrix field $\Phi (x)$ with  $D(x)\Phi D^\dagger(x)$, where $D(x)=
e^{ip_{\mu}x_{\mu}}$, and
$p_{\mu}$ is the diagonal matrix with matrix elements $p_{\mu}^{i}, i=1,...N$.
(quenching parameters). It has been shown that
quenched theory produces the standard Feynman diagrams for invariant
Green functions in all orders in perturbation theory.

Performing the reduction procedure $\Phi (x+\mu)=D_{\mu}\Phi
(x)D_{\mu}^\dagger$
and a change of variables $U_{\mu} \to U_{\mu}'=U_{\mu} D_{\mu}$ we left with
an action
\begin {equation} 
                                                          \label {2.1}
S(\Phi, U, D)=-\frac{1}{2}m^{2}N\tr \Phi \Phi^\dagger +\beta N \sum _{\mu >0}
\tr [\Phi U_{\mu}\Phi ^\dagger D_{\mu}^\dagger+
     \Phi U_{\mu}^\dagger \Phi ^\dagger D_{\mu}]
\end   {equation} 

Integrating over $\Phi$ and $\Phi ^\dagger$
one gets the sum over $\Gamma$ of  terms $m^{-2l(\Gamma)}
\tr U(\Gamma)\tr D(\Gamma))$.
Since all $D$  are diagonal matrices for the closed  contour $\Gamma $
we have $ D(\Gamma)=I$, and, therefore, we can neglect the last multipliers
in both terms
in equation (\ref {2.1}) and we left with the following reduced action
\begin {equation} 
                                                          \label {2.2}
S_{red}(\Phi, U)=-\frac{1}{2}m^{2}N\tr \Phi \Phi^\dagger +
\beta N \sum _{\mu >0}
\tr [\Phi U_{\mu}(x)\Phi ^\dagger +
     \Phi U_{\mu}^\dagger \Phi ^\dagger ]
\end   {equation} 

An other way of obtaining the reduced action (\ref {2.2}) consists in
application of
the reduced procedure directly in the induced action (\ref {1.5}). Performing
the reduction procedure for the gauge field  $U_{\mu}(x+\nu)=D_{\nu}U_{\mu}(x)
D_{\nu}^\dagger $ we get the  sum of  terms $U(\Gamma)$, this answer follows
also from  (\ref {2.2}).
The same action gives the approximation of the translation invariant
master field.

Hence, we have to estimate at large $N$ the following free energy
\begin {equation} 
                                                          \label {2.3}
F=\ln \int d\Phi d U_{\mu}\exp ((-\frac{\bar{m}^{2}}{2}N\tr \Phi
\Phi^\dagger +\bar{\beta } N \sum _{\mu >0}
\tr \Phi (U_{\mu}+U_{\mu}^\dagger )\Phi ^\dagger ))
\end   {equation} 
where $~\bar{\beta}=a^{d}\beta,
{}~\bar{m}^{2}=a^{d}m^{2}$.

Note that in (\ref {2.3}) $\Phi$ is an arbitrary complex $N\times N$ matrix.
Any matrix $\Phi$ may be written  in the form $\Phi=PV$ with $P$ hermitian
and $V$  unitary.
$P$ may be diagonalized by a unitary transformation, $P=S\Lambda S^\dagger $.
If we perform the change of variables $U\to S^\dagger VUV^\dagger S$ in the
integral
\begin {equation} 
                                                          \label {2.4}
I(\Phi ,\Phi ^\dagger )=\int d U_{\mu}\exp (\beta N \sum _{\mu >0}
\tr \Phi (U_{\mu}+U_{\mu}^\dagger )\Phi  )
\end   {equation} 
it is then clear that the integral (\ref {2.4}) depends only upon $\Lambda$.
Up to a normalization factor the measure  $d\Phi$ has the following form
$d ^{2N^{2}}\Phi =\Delta ^{2}(\lambda)\prod _{i}d\lambda _{i}dSdV$.
The integral (\ref {2.4}) defines the well known object,
this is  the partition function of the Brezin-Gross model which describes
 one link gauge field in the external matrix source \cite {BrN,BG}.
The form of this partition function in large N limit depends on the
magnitude of external source. In our case this parameter is
\begin {equation} 
                                                          \label {2.4'}
{\cal S}=\frac{1}{N\bar{\beta} }\tr (\Phi \Phi ^\dagger)^{-1} =
\frac{1}{N\bar{\beta}}\sum _{i} |\phi _{i}|^{-2},
\end   {equation} 
where $\phi _{i}=(\Lambda )_{ii}$. The weak coupling regime corresponds
to ${\cal S}<2$, and the strong coupling
regime to ${\cal S}>2$. The answer for both regimes can be written in
the following form  \cite {BG}
\begin {equation} 
                                                          \label {2.5}
\frac{{\cal F}}{N^{2}} = \frac{2}{N}\sum (\bar{\beta} ^{2}|\phi|^{4}+c)^{1/2}-
\frac{1}{2N^{2}}\sum _{i,j}
\log [(\bar{\beta} ^{2}|\phi _{i}|^{4}+c)^{1/2}+(\bar{\beta} ^{2}|\phi
_{j}|^{4}+c)^{1/2}]
-(c+3/4)
\end   {equation} 
The constant $c$ is different for different regimes. For the weak coupling
regime $c=0$  \cite {BrN} and for the strong coupling regime the constant
is determined from the equation
\begin {equation} 
                                                          \label {2.6}
\frac{1}{N} \sum (\bar{\beta} ^{2}|\phi _{i}|^{4}+c)^{-1/2}=2
\end   {equation} 
The next step is the calculation the integral in respect of $\phi$'s
\begin {equation} 
                                                          \label {2.7}
F=\ln \int \prod _{i,j}(\phi _{i}-\phi _{j})^{2} \prod d\phi _{i} dc
 [\theta ({\cal S}-2)\delta(c)+\theta (2-{\cal
S})\delta(c-\frac{1}{N\bar{\beta}}
\sum _{i} |\phi _{i}|^{-2})]\cdot
\end   {equation} 
$$
\exp [-\frac{\bar{m}^{2}}{2}N\sum \phi ^{2}_{i}+
2Nd\sum (\bar{\beta} ^{2}|\phi|^{4}+c)^{1/2}-$$
$$
\frac{d}{2}\sum _{i,j}
\log [(\bar{\beta} ^{2}|\phi _{i}|^{4}+c)^{1/2}+(\bar{\beta} ^{2}|\phi
_{j}|^{4}+c)^{1/2}]
-(c+3/4)]$$
\subsection   {Week Coupling Reduced Model}
One can roughly estimate the behaviour of this model in week coupling regime
ignoring the contribution
from the second term in the square brackets in the first line of (\ref {2.7})
and substituting  in (\ref {2.7})  an average of
\begin {equation} 
                                                          \label {2.8}
<\phi _{j}^{2}>=\frac{1}{N}\tr \Phi ^{2}=\frac{1}{\mu}.
\end   {equation} 
instead of the sum over $j$.  In this case we get the usual 1-matrix problem
in the logarithmic potential
\begin {equation} 
                                                          \label {2.8a}
{\cal V}=-\frac{m_{0}^{2}}{2}N\tr \Phi ^{2}+\frac{d}{2}\tr\log (1+\mu\Phi ^{2})
\end   {equation} 
where $m_{0}^{2}=4d\bar{\beta}-\bar{m}^{2}$. The parameter $\mu $ should be
defined from the selfconsistent equation (\ref {2.8}).
Taking into account only quadratic terms in (\ref {2.8a}) we get the following
selfconsistent equation
\begin {equation} 
                                                          \label {2.9}
d\mu -m_{0}^{2} =\mu ,
\end   {equation} 
which has a solution  for $d>1$ only if $m_{0}^{2}>0$, i.e
$\bar{\beta}_{\underline{cr}}=\bar{m}^{2}/4d$ is a critical point (the symbol
$\underline{cr}$ means that this critical point is obtained starting from week
coupling regime).  There is also a restriction
on $d <5/4$ coming from the requirement of applicability of the used
approximation.  According to
this requirement the region $|\phi|<2/\sqrt{\mu}$ must be inside the region
where
the potential (\ref {2.8a}) increases, i.e. $4/\mu < d/m_{0}^{2}-1/\mu$.

However there is a difficulty which makes the application of the
quadratic potential $V(\Phi)$ in the weak coupling  regime non-consistent.
It is caused by a breaking of the equality
${\cal S}<2$. This gives a raison to add a quartic term to the potential
\begin {equation} 
                                                          \label {2.10}
V(\Phi)=\frac{\bar{m}^{2}}{2}N\tr \Phi ^{2}+\bar{g}N\tr \Phi ^{4}
\end   {equation} 
Here $\bar{m}^{2}<4d\bar{\beta}$, and, therefore,  $m_{0}^{2}$ is
positive, i.e. we have the Goldstone potential; $\bar{g}=a^{d}g$. In this case
equation  (\ref {3.4}) gives
$\mu = \frac{2m_{0}^{2}}{2d+1}$. In the adopted approximation  the condition
${\cal S}<2$ means $ \frac{\bar {\beta} m_{0}^{2}}{g_{eff}}>2$, that can be
achieved
for $\beta$ being large enough; $g_{eff}=\bar{g}-\frac{d}{4}\mu $.
This approximation has also the  following restriction
$\frac{\bar {\beta} m_{0}^{2}}{g_{eff}}>\frac{8(1+2d)}{m_{0}^{2}}$. Note that
all the numerical factors are obtained in the approximation
of the quadratic potential in the vicinity of the minimum of the Goldstone
potential.

One should note that the application of the reduction procedure itself for the
weak
coupling regime is not quit correct since in this region as it was mentioned in
the
Introduction one has to use the reduction with constrains (see  \cite {IA92}).
However the simple calculations
presented above show that we have to deal with the effective Goldstone
potential and this gives an indication that  the structure of the model with
quenching momentum
may be  similar to the phase structure   of spin glasses  \cite {IA82}.

\subsection   {Strong Coupling Reduced Model}

Now we are going to estimate the free energy (\ref {3.2}) in the strong
coupling regime. In this case the second term in the first line of (\ref {3.5})
gives a main contribution, i.e. we have to deal with (\ref {3.3}), where
$c$ satisfies (\ref {3.4}). Taking the  following approximation for (\ref
{3.3})
\begin {equation} 
                                                          \label {2.11}
{\cal F} =-Ng_{eff}\beta ^{2}\sum \phi ^{4} +const, ~~
g_{eff}=\frac{1+2\sqrt{c}}{2\sqrt{c}(1+\sqrt{c})}
\end   {equation} 
we get the one matrix problem in quartic potential with a wrong sign. A phase
transition  in this model occurs at $\bar{\beta} _{\bar{cr}}=\frac{1}{8}
\sqrt{\frac{5}{3}}\frac{m^{2}}{\sqrt{d}}$, the point where
the function of distribution of eigenvalues lose the positivity (here
$\bar{cr}$
means that this critical point is obtained starting from strong coupling
regime).
It is interesting to compare the strong coupling regime of (\ref {2.3}) with
the strong coupling regime of the KM model. We will do it
in the forthcoming publication.

The above estimations advocate for the following picture. There are two
phase in the theory (\ref {2.3}), weak and strong phases. They are separated
by the intermediate region $(\bar{\beta} _{\bar{cr}},\bar{\beta}
_{\underline{cr}})$
.  The presented estimations are rather rough and in principle they can be made
more
precise if one does not restrict by the quartic potential and deal with
logarithmic
potential. However we believe  that the qualitative picture of the phase
structure
of the model (\ref {2.3})  is correctly reproduced. The condition $\bar{\beta}
_{\bar{cr}}<\bar{\beta} _{\underline{cr}}$  gives a restriction
on $d$.

\section   {Reduced Model with Quenching }
In the case of gauge theories  the quenching of the momentum must be
accompanied by a constraint on the eigenvalues of the covariant derivative.
Without this constraint one gets naive reduced model  without quenching. In the
case of the Wilson gauge theory the naive reduce model describes correctly
the theory only in the strong coupling regime.
The suitable constraint restricts
the eigenvalues of $U_{\mu}D_{\mu}$ to be equal to $D_{\mu}$.
This means that we have to take the measure
$d\mu (U)$ in (\ref {2.3}) to be
\begin {equation} 
                                                          \label {3.1}
d\mu (U)=\prod dU_{\mu} C(U_{\mu},D_{\mu})
\end   {equation} 
where $dU_{\mu}$ is the Haar measure on $SU(N)$  and
\begin {equation} 
                                                          \label {3.2}
C(U_{\mu},D_{\mu}) =\prod _{\mu}\int dV_{\mu}~
\delta(U_{\mu}-V_{\mu}D_{\mu}V_{\mu}^\dagger D_{\mu}^\dagger )
\end   {equation} 
If we integrate out the $U_{\mu}$, setting
$U_{\mu}=V_{\mu}D_{\mu}V_{\mu}^\dagger D_{\mu}^\dagger $
we obtain
\begin {equation} 
                                                          \label {3.3}
\exp(\tilde{F}(p)) =\end   {equation}
$$\int d\Phi \prod _{\mu} d V_{\mu}
\exp \{a^{d}(-\frac{1}{2}m^{2}N\tr \Phi \Phi^\dagger +\beta N \sum _{\mu >0}
\tr [\Phi V_{\mu}D_{\mu}V_{\mu}^\dagger \Phi ^\dagger D_{\mu}^\dagger+
     \Phi V_{\mu}D_{\mu}^\dagger V_{\mu}^\dagger \Phi ^\dagger D_{\mu})]\}
$$
To find the vacuum energy at the large $N$ limit one should integrate
$\tilde{F}(p)$ over $dp_{\mu}$. In fact we will use the formula
\begin {equation} 
                                                          \label {3.4}
F=\int d\mu (p)\tilde{F}(p),
\end   {equation} 
where
\begin {equation} 
                                                          \label {3.5}
d\mu (p)=\prod _{i}\frac{d^{d}p_{i}}{(\frac{\sqrt{\pi}}{a})^{d}}
\exp (-p^{2}_{i}a^{2}) ,
\end   {equation} 

So, the reduced action with quenching for the model (\ref {1.2})
contains the quadratic dependence from the unitary matrix $U$.
Note that  the reduced action with quenching for the KM model has the
quartic dependence from the unitary matrix, that makes the model not exactly
solvable.

The first step in the calculation of (\ref {3.3}) is the  evaluation of
the integral over $V_{\mu}$ 's. It is interesting
to note that just for this case the technics developed by Kazakov and Migdal
 \cite {KM,Mig92a}, and Gross  \cite {Gross} permits to study analytically the
problem in question.

Using the two-matrix representation of the Itzykson-Zuber (IZ) integral
\cite {IZ} and the orthogonal polynomial method  \cite {Metha} the following
integral over $V_{\mu}$ 's
\begin {equation} 
                                                          \label {3.7}
\exp I(\{\phi ^{2} \}, \{p_{i}\})=\int  d V
\exp (N \tr \Phi (V (D +D ^\dagger )V ^\dagger
\Phi ^\dagger ))
\end   {equation} 
at large N can be represented as
\begin {equation} 
                                                          \label {3.8}
\exp I(\{\phi ^{2} \}, \{p_{i}\})=
N^{2}\int _{0}^{1}dt (1-t)\ln f(t),
\end   {equation} 
where
\begin {equation} 
                                                          \label {3.9}
f(t)= -t+\oint \frac{dz}{2\pi i} \frac{V'_{1}(p(z),t)}{z^{2}},
\end   {equation} 
$$p(y,t)=\frac{1}{y}+\oint \frac{dz}{2\pi i} \frac{V'_{2}(p(z),t)}{1-zyf(t)}$$
$$q(y,t)=\frac{1}{y}+\oint \frac{dz}{2\pi i} \frac{V'_{1}(p(z),t)}{1-zyf(t)}$$
and
\begin {equation} 
                                                          \label {3.10}
V_{1}(x)=\frac{1}{N}\sum _{b}\ln (x-\phi ^{2}_{b}),
V_{2}(y)=\frac{1}{N}\sum _{b}\ln( y-2\cos ap_{b}),
\end   {equation} 

 We left only linear dependence on $D$
and $D^\dagger $ in the action in (\ref {3.7}), since the dropped $D$'s
do not contribute to closed loops. Note that in spite of  this reason
such step needs more arguments.

The next step is the evaluation of the integral over $\Phi$  and $p_{\mu}$.
Using the replica trick  \cite {Parisi,IA82}
$$\ln \int d\phi \Delta ^{2}(\phi )
\exp \{I(\{\phi ^{2}\},\{p_{i}\})\}=
$$
\begin {equation} 
                                                          \label {3.11}
\lim _{n\to 0}\frac{1}{n}
\left( \int dp\prod _{\alpha =1}
^{n} d\phi ^{\alpha}\Delta ^{2}(\phi ^{\alpha})
\exp \{I(\{(\phi ^{\alpha })^{2}\},\{p_{i}\})\}-1\right)
\end   {equation} 
one can perform first the integration over $p_{\mu}$'s
and then use the saddle-point approximation for integral over $\phi$
\begin {equation} 
                                                          \label {3.11a}
2\sum _{j\neq i}\frac{1}{\phi ^{\alpha }_{i}-\phi ^{\alpha }_{j}}=
2N\frac{m^{2}_{0}}{\beta}\phi ^{\alpha }_{i}-
d\frac{\partial }{\partial \phi ^{\alpha }_{i}} {\cal J}(\{\phi ^{\alpha }\})
\end   {equation} 
where
\begin {equation} 
                                                          \label {3.12}
\ln {\cal J}(\phi ^{\alpha })=\int dp \exp I(\{(\phi ^{\alpha })^{2} \},
\{p_{i}\}),
\end   {equation} 
$\alpha $ is the  replica index. Equation (\ref {3.11a}) is  more
complicated then the corresponding equation for the reduced KM model and
at present we do not know how to solve it. However,
an obvious advantage  of the model (\ref {1.1})  is that the IZ integral
completely describes the behaviour of the model,
while this  integral describes the KM model only in the approximation
without quenching, and even exhaustive knowledge of the
result of integration of the IZ integral over eigenvalues
of $\Phi$ does not provide information about  the behaviour of the KM
model in all regions of coupling constant.

\section {Concluding Remarks}
A lattice gauge model with simple linear interaction between gauge and
scalar fields was considered. This model does not suffer from local $U(1)$
symmetries  and it contains the Wilson action as the first term in the
expansion of the effective action. So, it seems the model can be considered
as a rather suitable approximation to QCD. It was shown that the model
in the large N limit  admits the analytical investigation and has the rich
phase structure. There are many open problems with the model. The most
important question here, as well as for the KM model, is a clarification
of the status of asymptotic freedom.

There are also some more technical problems which must be trieted out. First,
it would be interesting to study the reduced model in an external random field.
The action of this model is given by the expansion (\ref {2.1}).
Note that the role of the random field $D$ consists in the suppression of
contributions coming from open loops. A similar suppression may be ensured
by  abelian random fields $\exp (i\varphi _{\mu})$. Such model will be a
subject of our forthcoming
paper  \cite {IAp}. Another technical questions, as it has  been mentioned in
Section3, are related with more thorough estimation of the free action
(\ref {2.7}).

In concluding let us note once again that though there are many open
questions with induced QCD, this subject seems to be very promising
and undoubtedly  is worthy of further developments.

\end{document}